\documentclass[aps,prl,reprint,superscriptaddress]{revtex4-2}
\usepackage[utf8]{inputenc}
\usepackage{bm}
\usepackage{graphicx}
\graphicspath{{Pictures/}}
\usepackage{amssymb} 
\usepackage{amsmath} 
\usepackage{braket} 
\usepackage{natbib} 
\usepackage[colorlinks = true, linkcolor = magenta, urlcolor  = purple, citecolor = teal, anchorcolor = blue]{hyperref}
\usepackage{xcolor}
\usepackage{MnSymbol}
\usepackage[T1]{fontenc}
\usepackage{multirow}
\usepackage{tabularx}

\newcommand{\be}{\begin{equation}}
\newcommand{\ee}{\end{equation}}
\newcommand{\bea}{\begin{eqnarray}}
\newcommand{\eea}{\end{eqnarray}}

\begin{document}

\title{Exact Hopfion Vortices in a  3D Heisenberg Ferromagnet}
\author{Radha Balakrishnan}
\email{radha@imsc.res.in} 
\affiliation{The Institute of Mathematical Sciences, Chennai 600 113, India } 
\author{Rossen Dandoloff}
\email{rdandoloff@yahoo.com}
\affiliation{Department of Condensed Matter Physics and Microelectronics, Faculty of Physics, 
Sofia University, 5 Blvd. J. Bourchier, 1164 Sofia, Bulgaria} 
\author{Avadh Saxena}
\email{avadh@lanl.gov}
\affiliation{Theoretical Division and Center for Nonlinear Studies, Los Alamos National Laboratory, Los Alamos, New Mexico 87545, USA}

\date{\today}

\begin{abstract}
We find  exact static soliton solutions for the unit spin vector field of an 
inhomogeneous, anisotropic three-dimensional Heisenberg ferromagnet. Each 
soliton is labeled by two integers $n$ and $m$. It is a (modified) skyrmion in the $z=0$ 
plane with winding number $n$, which  twists out of the plane $m$ times in the $z$-direction to become a 3D soliton.   
Here $m$  arises due to  the  periodic boundary condition  at the $z$-boundaries.
We use Whitehead's integral expression  to 
find that  the  Hopf invariant  of the 
soliton is an integer $H =nm$. It  represents a 
hopfion vortex. 
Plots of the preimages of this topological soliton show that 
they are either unknots or 
nontrivial knots, depending on $n$ and $m$. 
Any pair of preimage curves 
links $H$ times, corroborating  the 
 interpretation of  
$H$ as a linking number.
We also calculate the exact energy of the
hopfion vortex, and show that its topological lower 
bound  has a sublinear dependence on $H$. Using Derrick's scaling analysis, we  demonstrate that the presence of a spatial inhomogeneity in the anisotropic interaction, which in turn introduces a characteristic length scale in the system,  leads to  the stability of the hopfion vortex.
\end{abstract}

\maketitle

\textit{Introduction.} Three dimensional (3D) topological  solitons
are of great current interest. They have recently been observed in magnetic \cite{kent}, ferrroelectric \cite{ferroelectric}, liquid crystal \cite{LC1, LC2}, and other materials as well as in photonics \cite{photonics}, and studied in Bose-Einstein condensates \cite{BEC1, BEC2}. 
As is well known, solitons \cite{rajaraman}--\cite{shnir} are spatially localized, particle-like excitations that arise as  solutions of nonlinear partial differential equations satisfied by  the 
field configurations of the physical system concerned. 
A soliton can be non-topological or  topological.  Unlike the former, the latter is endowed with a nontrivial integer topological invariant, also called its topological charge. 
This topological property of the entity along with its energetic stability, is expected to become useful in communication technology, since  these particle-like  nonlinear topological excitations can serve as information carriers \cite{photonics}.

Various types of Heisenberg exchange models for interacting spins  describing  a number of magnetic materials are  storehouses of solitons \cite{kosevich}.   In the static case of the classical continuum version,  a normalized spin  configuration at any point 
$\mathbf{r}$ in physical space is described 
 by a {\it unit}  vector field  $\mathbf{S}(\mathbf{r})$.  Clearly, the tip of such a spin vector   
 lies on a $2$-sphere  $S^2$,  irrespective of the spatial dimension in which it  exists. 
In 2D,  the topological solitons are the well known 
magnetic skyrmions. These are classified by  an integer topological invariant (Pontryagin charge) 
$Q = (1/4 \pi)\iint \mathbf{S} \cdot
 (\partial_{x}\mathbf{S} \times \partial_{y}\mathbf{S})\,  dx \, dy$, 
 called the winding number \cite{rajaraman}, characterizing the second homotopy group $\pi_{2}(S^2) =Z$.
First studied  in 1975 by Belavin and Polyakov \cite{belavin}  in the context of 2D  isotropic ferromagnets, they  have  been investigated 
theoretically 
in other  magnetic models by several  authors. They have  also been observed  experimentally in many types of  2D magnetic materials \cite{reviews}. The  possible role of  magnetic skyrmions as bits to store information in future computer technology  has been suggested \cite{reviews}.
 
 In 3D, such solitons are classified by a topological invariant called the Hopf invariant $H$ (or Hopf charge), which is given by the Whitehead integral expression \cite{whitehead} [see Eq. (\ref{hopf-inv}) below]. 
 Here $H$ can also be  interpreted as the linking number of the  two closed space curves in 3D physical space that are the    preimages of any two distinct points on the target space $S^{2}$. 
  Magnetic materials provide an ideal platform to create and study 
  such topological solitons experimentally \cite{kent}.  Their  investigation  as   possible static solitons in  3D Heisenberg  models  is  therefore of current interest. They arise as  solutions of  the  variational 
 equations minimizing the energy, 
 the latter generically being 
  nonlinear partial differential equations that are  difficult to  solve analytically.   Hence existing theoretical work on topological solitons  typically uses numerical methods as well as simulations 
  \cite{sutcliffeFrust}--\cite{wang}. These studies have undoubtedly yielded useful insights regarding 3D spin textures as well as the knots and links associated with them.

 In the case of most micromagnetic models such as \cite{rybakov2},  the use of   numerical techniques  is unavoidable. On the other hand, it  is  instructive to identify a physically realizable magnetic model in 3D in  which  both the exact  soliton
solution as well as its  corresponding  Hopf invariant can be calculated  {\it analytically}.  Analytical methods  play a crucial  role in clarifying the  basic  physical and topological 
characteristics of  solitons.  Recently,  topological solitons have been created and observed experimentally in a multilayer magnetic system \cite{kent}. A solvable model  can  also suggest the fabrication of appropriate magnetic materials and  initiate more experiments  to study the various topological aspects of these nonlinear excitations. The present work is motivated by  
these  considerations.
  
  Our main results are as follows:  We   find  
   exact  static soliton solutions for the unit spin configurations 
   $\mathbf{S}(\mathbf{r})$ 
   of  a 3D,  {\em inhomogeneous},  anisotropic Heisenberg 
   ferromagnet. Each soliton is labeled by two integers $n$ and $m$. It is a  modified skyrmion  in the $z=0$ plane with winding number $n$, which  twists out of the plane  to become a 3D soliton. Here $m$  arises from  the  periodic boundary condition  imposed in the $z$-direction. 
    Using the Whitehead formula \cite{whitehead}, we calculate its Hopf charge  analytically  to obtain  an  integer $H =nm$. It represents  a hopfion vortex. 
   ($H < 0$ corresponds to a hopfion antivortex.) Using the exact  solution, we plot the preimages of a few distinct points on a specific latitude of the target space $S^2$, and show that they are closed space curves that lie on a corresponding $2$-torus. [$\mathbf{S}(\mathbf{r})$ 
 points in a fixed  direction on a preimage curve.]    
   These curves are either unknots or nontrivial knots, depending on $n$ and $m$. Any two of them link $nm$ times, yielding the geometric  interpretation of $H$ as a linking number.  Thus, this hopfion vortex is associated with a twisted, knotted, linked structure. The preimages of 
 the points on any  latitude  of $S^{2}$ densely  fill the surface of 
 an associated  torus. We  then calculate the exact  energy $E$ of the magnetic hopfion vortex. We further  find  that  $E  \geqslant   c H^{1/2}$ where $c$ is a material dependent constant, 
 showing  that  
the  topological lower bound on $E$ has a  {\it sublinear}  dependence on the Hopf charge. Using Derrick's scaling analysis \cite{hobart, derrick}, we  show that the presence of the  spatial inhomogeneity in the anisotropic interaction, which 
in turn introduces a characteristic length scale in the model, leads to  the stability of the hopfion vortex.

\textit{Exact solitons for a 3D Heisenberg model.}  
  We consider the continuum version of a magnetic system described by a  classical anisotropic ($XXZ$), {\em inhomogeneous}  Heisenberg ferromagnet, with energy $E$ given by
\be 
E =  (J/a)\, \iiint \big\{  (\partial_{x}\mathbf {S})^2 + 
(\partial_{y}\mathbf {S})^2 
+ \widetilde{J}_{3}(\rho)
(\partial_{z}\mathbf {S})^2 \big\} 
\,dx \,dy \,dz.
\label{Hc}
\ee 
Here $J$ is  the nearest-neighbor 
exchange interaction   in the $x$ and $y$ directions,
 $\widetilde{J}_{3}(\rho) = J_{3}(\rho)/J$ is the dimensionless, {\em inhomogeneous}  anisotropic  interaction in the $z$-direction, with $\rho =\sqrt{x^2+y^2}$, and $a$ is the lattice constant. 
 
 In what follows, we will show that an  {\em inhomogeneous}  anisotropy of the form  $\widetilde{J}_{3} (\rho)
= K_3 \,l^{2}/\rho^2$ in Eq. (\ref{Hc})
leads to {\it exact solutions}  for the spin textures $\mathbf{S}(\mathbf{r})$.  Here,  
$K_{3}$ is the strength of the anisotropy and  $l$ is the length  scale characterizing the inhomogeneity. In addition, this   functional form also ensures the stability of the exact spin textures obtained, as will be  explained  in detail later.

The unit vector   $\mathbf{S}$  is given in  spherical polar 
coordinates by 
\be
\mathbf{S} = (\sin\Theta\, \cos \Phi,\, \sin\Theta \, \sin \Phi, \,\cos\Theta).
\label{S}
\ee
Substituting this  in Eq. (\ref{Hc}) and  transforming
to  cylindrical coordinates $(\rho, \phi, z)$ in physical 
space, we get 
\bea 
E &=& (J/a) \iiint\!
\Big\{ 
\big[
(\partial_{\rho}\Theta)^{2}+ \rho^{-2}
(\partial_{\phi}\Theta)^{2}\big] \nonumber\\ 
&+& \sin^2\Theta 
\big[
(\partial_{\rho}\Phi)^{2}+ \rho^{-2}
(\partial_{\phi}\Phi)^{2}\big]
 \nonumber \\ 
&+& \widetilde{J}_{3}(\rho)  \big[
(\partial_{z}\Theta)^{2}
+ \sin^2\Theta 
\,(\partial_{z}\Phi)^{2}
\big]   \Big\} \rho~ d\rho \,d\phi \,dz.
\label{Hc1}
\eea 
We consider solutions of the form $
\Theta = \Theta(\rho), \, \Phi = \alpha_0 \phi + \beta_0 z +\Phi_{0}$, 
where the constants $\alpha_{0}$ and $\beta_{0}$ 
are to be determined by the boundary conditions on  $\Phi$.
Equation  (\ref{Hc1}) then reduces to 
\bea 
E &=&  (J/a)\, \iiint \big\{ 
\rho \,(\partial_{\rho}\Theta)^2 \nonumber\\ 
&+&\alpha_0^{2}\,\rho^{-1}\,\sin^2\Theta 
+ \widetilde{J}_3(\rho) \,\beta_0^2 \,\rho \,\sin^2\Theta \big\} 
d\rho \,d\phi \,dz   \,. 
\label{Hc2}
\eea 
Setting  $\widetilde{J}_{3}(\rho) = K_3 \,l^{2}/\rho^2$ [mentioned below Eq. (\ref{Hc})],
we find the Euler-Lagrange  equation 
for the energy functional in Eq. (\ref{Hc2}). Then,
 changing  variables to  $\widetilde{\rho} = 
 \ln \,(\rho/\rho_{0})$ \cite{skyrmion}  (where $\rho_0$ is a  constant)  in this equation, we obtain
 \be
 \partial^{2}\Theta/\partial \widetilde{\rho}^{2} =  
 \tfrac{1}{2} \left(\alpha_0^2 + K_{3} \,l^{2}\,\beta_0^2\right) \sin2\Theta \,.
\label{homo2}
\ee 
 Imposing  the periodic boundary conditions 
$\Phi(\phi+2\pi) = \Phi(\phi)$ and  $\Phi(z+L) = \Phi(z)$, we find $\alpha_0 = n, \,\,\beta_0 = 2\pi m/L$
where $m$ and $n$ are integers. Here $L$  is a constant 
representing the thickness of the  given 3D magnetic system. 
Equation (\ref{homo2}) then yields, for the 
function $\widetilde{\Theta} = 2\Theta$,  
\bea 
\partial^{2}\widetilde{\Theta}/\partial\tilde{\rho}^2 = \mu^2 \sin\widetilde{\Theta} 
  \label{thetaBar}
\eea
with  the solution $\widetilde{\Theta}(\widetilde{\rho}) = 4 \tan^{-1}\left(e^{\mu \widetilde{\rho}} \right) $,
where 
\be
\mu = 
\pm  \big(n^2 + 
4K_{3}\pi^{2} m^{2}l^{2}/L^{2}\big)^{1/2}.
\label{mu}
\ee
In terms of the  original variables,
the solution for $\Theta$ reads
\be
\Theta(\rho) = 2 \tan^{-1} [({\rho}/\rho_{0})^{\mu}].
\label{tan}
\ee
Using the above solution in Eq. (\ref{S}),  we arrive at the following
exact static solution for the spin 
configuration in 3D analytically.
\bea 
\mathbf{S} (\rho, \phi, z) &=& \Bigg( \frac{2(\rho/\rho_{0})^{\mu}}{1+(\rho/\rho_{0})^{2\mu}} 
\cos\,\Phi(\phi, z),  \nonumber\\ 
&&\frac{2(\rho/\rho_{0})^{\mu}}{1+(\rho/\rho_{0})^{2 \mu}} 
\sin\,\Phi(\phi,z), 
 \frac{1-(\rho/\rho_{0})^{2\mu}}{1+(\rho/\rho_{0})^{2\mu}}    \Bigg), ~~
\label{hopfion}
\eea 
where  $\Phi(\phi,z) = n\phi + 2\pi m z/L +\Phi_{0}$, with $\Phi_0=\Phi(0,0)$.

Clearly, the possible spin configurations $\mathbf{S} (\rho, \phi, z)$ given in Eq. (\ref{hopfion}) 
are labeled by  two integers $n$ and $m$. It is important to note that  in the solution (\ref{hopfion}), $\mu$ is defined in Eq. (\ref{mu}), where  $K_{3}, l$ and $L$  are material parameters of our model.

If $\mu < 0$, then as $\rho\rightarrow0$, we find 
$\Theta(\rho) \rightarrow \pi$ and  
 $\mathbf{S}\rightarrow (0,0,-1)$; while 
 as $\rho \rightarrow \infty$, we have 
$\Theta(\rho) \rightarrow 0$ and hence  
   $\mathbf{S} \rightarrow (0,0, 1)$.   
 When  $\rho = \rho_{0}$, $\Theta = 
 \pi/2$.  
In the plane   $z=0$, the solution becomes a 
{\em modified} skyrmion (resp., antiskyrmion) 
for $n > 0$ (resp., $n < 0$).  (The modification 
arises essentially from the presence of the anisotropy 
$K_{3}$ with its inhomogeneity characterized by the length scale $l$, in the exponent $\mu$.)
 Its winding number (topological charge) $Q$  can be computed, to obtain   $-n$ and $+n$, respectively \cite{koshibae}. An inspection of  Eq. (\ref{hopfion}) shows that  this  skyrmion twists out into the $z$ direction in a periodic fashion $m$ times. Thus it is a  3D soliton describing a {\em twisted skyrmion string}.
 Such a solution has been found {\em numerically}  in the context  of  other magnetic models 
 \cite{yokota,nitta}.  
 
 The occurrence of  $\mu$ as the exponent of $(\rho/\rho_{0})$ in the soliton solution (\ref{hopfion}) is of significance. For a fixed $L$, the form and geometry of the topological solution we have obtained depend  on 
the  physical parameters $K_{3}$ and $l$
that appear in $\mu$, representing respectively the 
effects of anisotropy and inhomogeneity in the 
interacting system of spins. 
The presence of $\mu$ enables us to control the rate of change of 
$\Theta(\rho)$ with $\rho$ in the soliton solution, by  tuning  these material parameters. This in turn should be helpful in  designing experiments to create and observe the twisted 3D soliton. Usually, in a given experiment  it is convenient to keep $K_3$ and $L$ fixed, and examine the 3D spin textures for various  length scales  $l$ of the inhomogeneity. Indeed, 
one way to change $l$  systematically is to  vary the (functionally graded) doping profile in the $(x,y)$ plane appropriately, in experiments.

For completeness, we point out that if  $\mu > 0$ in Eq. (\ref{hopfion}), the spin configuration  for $z =0$ corresponds to $\mathbf{S} \rightarrow (0,0,1)$ as  $\rho\rightarrow 0$, while $\mathbf{S} \rightarrow 
(0,0,-1)$ as  $\rho \rightarrow \infty$. Some authors \cite{reviews} use this  alternative  boundary condition 
to define a skyrmion.  
All the  results in the foregoing discussion hold good for both conventions.  

\textit{Calculation of the Hopf invariant $H$.} 
 As mentioned in the Introduction, $H$  can be calculated  
  from the  Whitehead formula \cite{whitehead,gladikowski}
 \be 
 H = -(1/ 8\pi^{2}) \iiint (\mathbf{A} \cdot \mathbf{B}) \,
 dx\,  dy\, dz,
 \label{hopf-inv}
\ee
where the Cartesian components of the {\it emergent magnetic field} \cite{kent}  
are given by 
$B_{x}  = -\mathbf{S} \cdot(\partial_{y}\mathbf{S}\times\partial_{z}\mathbf{S})$ and cyclic permutations for 
$B_{y}$ and $B_{z}$, and $\mathbf{A}$ is the corresponding vector 
potential. 
 It is easily  verified  that $\nabla. \mathbf{B} =0$.  Using 
 the solution (\ref{hopfion}) and expressing the Cartesian 
 components of $\mathbf{B}$ in cylindrical polar coordinates in 
 physical space, we get 
\bea
 && B_x  =  \beta_0 
 ( \partial_{\rho} \cos \Theta)
 \sin\phi \,, ~~
 B_y =   - \beta_0 ( \partial_{\rho} \cos \Theta)
 \cos\phi \,,  \nonumber\\ 
&& B_z =  (\alpha_0/\rho)  
  \partial_{\rho} \cos \Theta.
 \label{b123}
 \eea 
 Solving 
  $\nabla \times \mathbf{A} =\mathbf{B}$ for the 
  Cartesian components of $\mathbf{A}$ 
using the appropriate boundary conditions on $\Theta (\rho)$ [as described below Eq. (\ref{hopfion})],  a lengthy but straightforward calculation yields 
\bea 
&&A_{x} = - (\alpha_0/\rho) (\cos\,\Theta \pm 1) \sin\,\phi,\, \nonumber\\  
&& A_{y} =  (\alpha_0/\rho)(\cos\,\Theta \pm 1) \cos\,\phi,\,  ~~
A_{z} = \beta_0 \cos\,\Theta .
\label{a123}
\eea 
The $\pm$ signs  correspond to $\mu > 0$  and $\mu < 0$, respectively.
Substituting Eq. (\ref{b123}) and Eq. (\ref{a123})  in Eq. (\ref{hopf-inv}),  the Hopf invariant   of the 3D soliton  can be written in the form
 \be 
 H =  \mp \frac{ \alpha_0\beta_0}{8\pi^{2}} \int_{0}^{L}\!dz \int_{0}^{2\pi}\!d\phi \!\int_{0}^{\pi} \sin\,\Theta \,  d\Theta.
 \label{H1} 
 \ee
 Since  $\alpha_{0}= n$ and
 $\beta_{0} = 2\pi m/L$,   
we obtain
\be
H =  nm,
\label{mn}
\ee
keeping in mind that $n$ can be a positive 
or negative integer. Interestingly, this integer $H$ 
emerges as a  product  of two integers
in our spin system. Note that both $m$ and $n$ have to be nonzero integers for the Hopf charge $H$ to be nonzero.

Usually, a 3D topological soliton is called a hopfion  if it satisfies uniform 
boundary conditions [e.g., ${\bf S}({\bf r} \rightarrow \infty)$ =$(0,0,1)$], so that the 3D physical 
space can be compactified to $S^3$. It represents a map $\mathbf{S}: S^3 \rightarrow S^2$. Its Hopf invariant  is  an integer  characterizing  the third homotopy group $\pi_{3}(S^2)$.   On the other hand, our soliton solution Eq. (\ref{hopfion}) described by a twisted skyrmion string  
is obtained using
the  homogeneous boundary condition  for
$\mathbf{S}$ in each   
$z = {\rm constant}$ plane, together 
with  the periodicity 
in the $z$-direction. This 
represents  a map $\mathbf{S}: S^2 \times T^1 \rightarrow S^2$ \cite{jaykka}.  Due to this difference, our twisted skyrmion string given in Eq. (9)  is  called a  {\it hopfion vortex}  rather than a hopfion. 
As seen from Eq. (\ref{mn}), its integer Hopf invariant is obtained as the product of the winding number $n$ of the skyrmion  in the $xy$ plane,  and  the integer $m$ giving the number of times it winds
 around the $z$-axis till it reaches the boundary at $z=L$.  These integers encode, respectively,  the topology of the $S^2$ and $T^1$  parts of  the manifold $S^2 \times T^1$. Since $H$ in Eq. (\ref{mn}) can have either sign, 
the system supports both hopfion vortices and 
 hopfion antivortices.

 \begin{figure}[h] 
\centering 
\includegraphics[width=2.0 in]{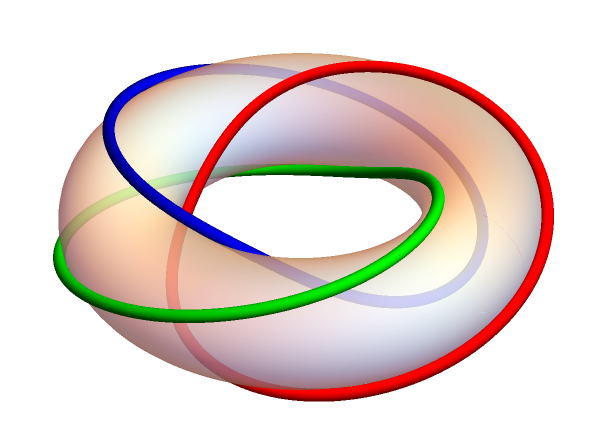} 
\includegraphics[width=2.0 in]{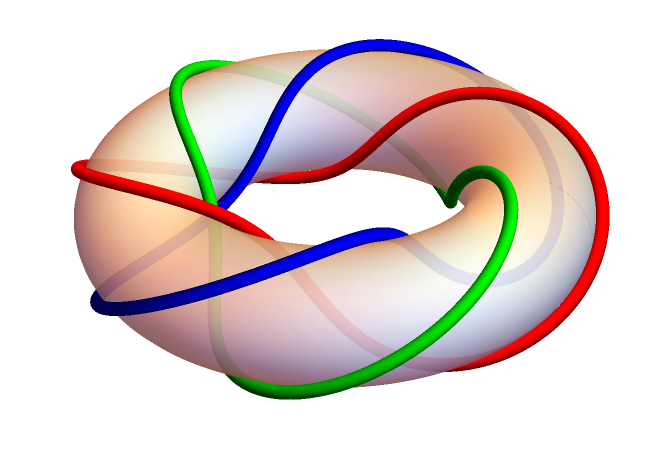} 
\includegraphics[width=2.0 in]{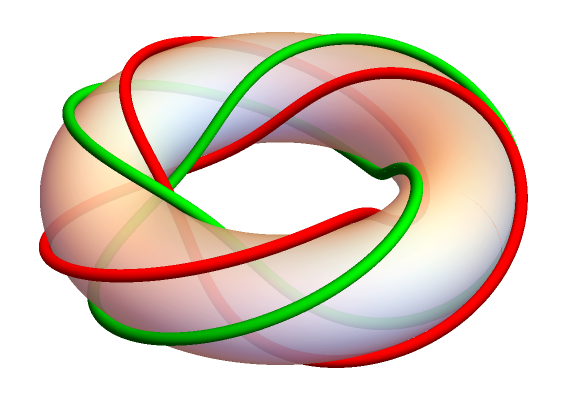}
\caption{Preimages on a torus for a hopfion vortex
[Eq. (\ref{hopfion})].
(a) Upper panel: $n=1, m=1$, unknots, linking 
number  $H = 1$. (b) Middle panel: $n=1, m=2$, unknots, 
linking number $H = 2$. 
(c) Lower panel: $n=2, m=3$, trefoil knots, linking number 
$H = 6$.}
\end{figure} 
 
\textit{Knotted structure of the hopfion vortex  and Hopf invariant as a linking number.}
Next, we use {\it  Mathematica} to find  the 
preimage   of any  specific point on $S^2$,
  i.e., the points in 3D space corresponding 
  to a specific value $(\Theta, \Phi)$ of  $\mathbf{S}(\mathbf{r})$ of the hopfion  vortex solution (\ref{hopfion}). 
   We have plotted the three  preimage curves corresponding to 
   $\Theta = \pi/2$   and $\Phi =0, \pi/3$ and $2\pi/3$, for the cases  (i) $n=1,m=1$ [Fig. 1(a)], 
   (ii) $n=1, m=2$ [Fig. 1(b)], and (iii) $n = 2, m = 3$ [Fig, 1(c)].
  As these illustrative examples show, each of the  preimages is a closed space curve which is, 
 in cases (i) and (ii),  
  an  unknot, topologically equivalent to a circle.
In contrast, it is 
     a trefoil knot (a nontrivial knot) in case (iii).   
 Further,  as can be readily seen in  Fig. 1, each closed curve  lies on the surface of a  torus,   traversing  $n$ times around the poloidal direction and $m$ times around  the toroidal direction.  The analytical result of Eq. (\ref{mn}) gives 
 $H = 1, 2, 6$, respectively in cases (i), (ii) and (iii). 
 Correspondingly, we see from Fig. 1 that
   any two closed space curves  link once, twice and six times, respectively, in these three cases. This corroborates 
   geometrically that the  Hopf invariant $H$  is 
   precisely just the {\it linking number}  of the preimages of two   distinct  points on $S^2$. 
  For a given value of $\Theta$, the preimages 
  of  the points $\Phi \in [0, 2\pi)$ densely fill the  
   corresponding torus (Fig. 2).
   As is well known \cite{oberti},  a torus knot  is an 
     unknot if and only if either $n$ or $m$ is $\pm1$,
     and  a  nontrivial knot if $m$ and $n$ are coprime.  
     Our plots illustrate  the knotted and linked 
     structure  of the hopfion vortex.  
  \begin{figure}[h] 
\centering 
\includegraphics[width=2.0 in]{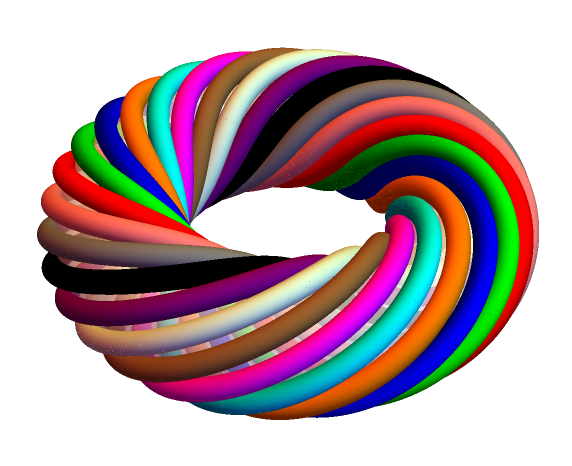} 
\caption{A torus densely filled by the preimages 
of the points on a fixed latitude of $S^2$. Different colors correspond to different 
points on the latitude.}
\end{figure}

\textit{Exact energy of the hopfion vortex  and its topological lower bound.}
Setting $\widetilde {J}_{3} = K_{3} l^{2}/\rho^{2}$
in Eq. (\ref{Hc2}), using the definition of 
$\mu$ from Eq. (\ref{mu}) and 
putting  in  the appropriate limits of integration, the  energy of the hopfion vortex is given by
\be
E = (J/a) \int_{0}^{L}\!dz\int_{0}^{2\pi}\!d\phi
\int_{0}^{\infty}\!
\big\{\rho (\partial_{\rho}\Theta)^2 + 
  \mu^{2}\rho^{-1}
  \sin^{2}\Theta\big\} \, d\rho \,. 
  \label{E}
  \ee  
Using Eq. (\ref{tan}) for $\Theta(\rho)$, 
a short calculation yields
\be
E =   \frac{16 \pi JL}{a} \,{\mu}^{2}  \rho_{0}^{2\mu}\, \int_{0}^{\infty} \, \frac {\rho^{(2\mu -1)}}{[ \rho_{0}^{2\mu} +\rho^{2\mu}] ^{2} } \, d\rho = (8 \pi J L/a)\, |\mu|.  
\label{energy1}
\ee
Substituting for $\mu$  from Eq. (\ref{mu}), 
the  energy of the hopfion vortex  is given by the exact expression 
\be
E(n,m) = (8 \pi J L/a) \big(n^2 + 
4 \pi^{2} K_{3} l^{2} m^{2}/L^{2} \big)^{1/2}.
\label{energy2}
\ee
Since $E(n,m) =E(-n,m)$,
 hopfion vortex and hopfion antivortex have the same energy.
 
 We write Eq. (\ref{energy2}) as
 \be
E(n,m) = (8 \pi J L/a) \big(n^2 + \beta^{2}\, m^{2}\big )^{1/2},
\label{energy3}
\ee
where we have defined
 \be
\beta = 2 \pi \sqrt{K_{3}} (l/L).
\label {beta}
\ee
Substituting the inequality 
$\big(n^2 + 
\beta^{2}m^{2} \big)  \geqslant 2 \beta  mn$
 in Eq. (\ref{energy3}), and using $mn =H$, we get 
\be
E(n,m) \geqslant  c \sqrt {H},
\label{ineq1}
\ee
where $c = 2^{4} \pi^{3/2} K_{3}^{1/4} J \sqrt{lL}/a$.
Thus, the lower bound of the energy of the hopfion vortex has a  {\it sublinear}  dependence on  its topological  charge $H$. This is in contrast to the well known  lower energy bound  for the skyrmion (a 2D topological soliton) which is linear in  the Pontryagin  charge $Q$.  Such a sublinear behavior is usually attributed \cite{ward}  to  the knotted and  linked preimages, which is the source of the charge $H$  of a 3D topological soliton \cite{fn}. 

\textit{Stability.} Before investigating the stability of our 3D hopfion vortex solution  given in Eq. (\ref{hopfion}), we  first carry out the {\em general}  Hobart-Derrick scaling analysis \cite{hobart, derrick} for the energy expression  $E$ given in Eq. (\ref{Hc}), after substituting the {\em inhomogeneous} anisotropy  $\widetilde{J}_{3}(\rho) = K_{3} l^{2}/\rho^{2}$  in it \cite{fn2}.

 It is convenient to  write the  first two terms  of  the energy [Eq. (\ref{Hc})] as
 \be
 (J/a)\, \iiint  [(\partial_{x}\mathbf {S})^2 + 
(\partial_{y}\mathbf {S})^2] dx\,dy\,dz = A_{0},
\label{A0}
\ee
 and its last term as
 \be
 (J/a)\, \iiint [K_3 \,l^{2}/(x^2+y^2)]\,
(\partial_{z}\mathbf {S})^2 \,dx\,dy\,dz =B_{0},
\label{B0}
\ee
where we have used $\rho^{2} = (x^{2} + y^{2})$.  Derrick's scaling analysis  \cite {derrick}   involves letting  $(x,y,z)$ to $(\lambda\, x,\lambda\, y,\lambda\, z)$  in  the energy expression of Eq. (\ref{Hc}), $\lambda$ being  the  scale factor.  This yields
\be
E(\lambda) = \lambda \,A_{0}\,+\, (1/\lambda)\, B_{0} \,. 
\label{Elambda}
\ee
 To analyze the extrema  of $E (\lambda)$, we set $dE/d \lambda = A_{0} -(1/\lambda^{2}) \,B_{0} \,=\,0$. This can have a  positive solution $\lambda_1 = +\sqrt{(B_{0}/A_{0})}$, where $A_{0}$ and $B_{0}$ are integrals  defined in Eqs. (\ref{A0}) and (\ref{B0}) above.  Note that $A_{0}$ is always positive and $B_{0}$ is positive for $K_{3} >0$, as considered in our model. Also note that $B_0$ has a  multiplicative factor that depends on $l^2$.  It can be easily verified that
$d^2E/d\lambda^2  = [2/(\lambda_{1})^3]  B_{0}$  is  {\em positive} for any finite value of $l$,  showing that $E(\lambda)$ has a minimum at  $\lambda_{1}$. This implies that there is {\em generic}  stability in this  {\em inhomogeneous} anisotropic  system.

Next, we demonstrate the  stability of the specific case of our hopfion vortex solutions  for ${\bf S}$ given in  Eq. (\ref{hopfion}), by computing  the  corresponding integrals $A_{0}$ and $B_{0}$  for these solutions explicitly. Our detailed (and somewhat lengthy) calculations  yield  
\be 
\lambda_{1}= \sqrt{B_{0}/A_{0}} = \beta m/\sqrt{[ 2n^2 + \beta^2 m^2]},
\label{lambda1}
\ee
where   $\beta$ is given in  Eq. (\ref{beta}).
It is readily seen  that  in experiments with a fixed  $K_3$ and $L$, for a suitable choice of
 $l$,  $\beta $ is  finite and fixed. Hence from Eq. (\ref{lambda1}),  the scale $\lambda_{1}$  is  seen to be finite for all  {\it nonzero} integers $m$ and $n$. This shows that  the  hopfion vortex  solutions  supported by the {\em inhomogeneous}  anisotropic  ferromagnet will  not  shrink or flatten out,  {\em establishing  their  stability}. 
 
 Note that  the  presence of the characteristic length $l$  of the inhomogeneity in the anisotropic term plays an important  role in the stability of 3D spin 
textures. This is reminiscent of several 2D models of spin systems  where the introduction of a  characteristic length in the system  via  diverse physical mechanisms  \cite{skyrmion, dandoloff}   typically leads to the stabilization of 2D spin textures.

We parenthetically remark  that a   {\em homogeneous}  anisotropy  corresponds to  setting $\widetilde{J}_{3}(\rho) =(J_{3}/J)$ in the second term in the energy [Eq. (\ref{Hc})]. It is easily verified that  Derrick's   scaling analysis  in this case will lead to $E(\lambda) = \lambda (\,A_{0}\,+\, B_{0} \,)$  [instead of Eq. (\ref{Elambda})],  showing  that there is no minimum value for  $E$ for any 
 $\lambda$. This confirms the well known result that  the  solutions of   a  {\em homogeneous}  anisotropic Heisenberg ferromagnet are generically unstable.

We also mention that our scaling analysis given above,  which proves the  stability of  solitons in a  3D magnet in the presence of an {\em inhomogeneous}  anisotropy,  is similar to the analysis usually given \cite{leonov} for proving the stability of  skyrmions in an anisotropic  2D magnet in the presence of a Dzyaloshinskii-Moriya interaction term.
It has  been shown in the case of 3D chiral ferromagnets \cite{liu}  and chiral ferromagnetic fluids \cite{smalyukh} that the presence of the Dzyaloshinskii-Moriya \cite{DM}   interaction term of the form   $D \,\mathbf{S}\cdot (\nabla \times \mathbf{S})$ in the energy plays an important role in  stabilizing  the soliton. Turning to nonchiral  (inversion symmetric)  3D ferromagnets,  it is reasonable to expect that continuum Heisenberg models with competing energy terms could  lead to stable solitons.  However, 
    identifying appropriate terms which would yield a  stable 3D soliton solution which also has  an integer Hopf invariant (as we have, in our model)  is far from obvious.

\textit{Discussion.}  The main results obtained in this paper have already been summarized in the Introduction. Our results are novel and we believe they open up new avenues of investigation, e.g. hopfion vortex {\em lattice}  solutions of the model, study of the effects of an applied magnetic field, topological transitions in spin textures, Berry phase phenomena and the dynamics of hopfion vortices. 

   The introduction of  an inhomogeneity in the exchange interaction  in a Heisenberg model was 
 motivated in part  by an earlier 
 work  \cite{balakrishnan2} on the  dynamics of the continuum model of  an isotropic  Heisenberg chain with an   inhomogeneous  exchange interaction, which supports stable 1D solitons for certain specific inhomogeneities.       
 Since then,  various  aspects of  inhomogeneous magnetic systems have been studied by several other authors \cite{inhom}.
  
    We  remark in passing that  the results we have
   presented for the continuum Heisenberg model  should be applicable in  fields other than magnetism, where the corresponding 
  energy density involves inhomogeneous, anisotropic generalizations 
  of  $|\nabla \mathbf{n}|^{2}$, where $\mathbf{n}$ is a unit vector field. The energy density of the nonlinear sigma model \cite{rajaraman}, the splay term in the free energy of liquid crystals \cite{LC2}, the curvature term in the elastic rod energy \cite{harland}, etc. are some examples.
      
 Theoretical and experimental studies of  topological solitons  in 3D Heisenberg models  have started to gain momentum in recent years. There is a recent numerical study \cite{yokota} on twisted skyrmions which become hopfion vortices for appropriate boundary conditions. Hopfions have been identified in chiral ferromagnetic fluids \cite{smalyukh} 
 and observed \cite{kent}  in magnetic multilayer systems. Based on  
 their nanometer to micrometer sizes in various
 magnetic materials as well as
   their topological  and energy-based stability, the  possible   application of topological solitons in future  computer technology has been recognized. 
   They can be used to store bits of information, where a bit  corresponds to the presence or absence of  a topological soliton.  Certain dynamical advantages of  3D localized entities over skyrmions as information carriers have also been pointed out \cite {wang, photonics}. Thus, one could envisage such distinct applications as {\it hopfionics} akin to the field of skyrmionics \cite{reviews}. 
 
  We conclude by pointing out that   our magnetic model is not just an exactly solvable theoretical model  that reveals all the topological aspects of the 3D topological solitons obtained by us succinctly, but is also useful in designing  novel experiments to observe them. Specifically, we note  that  the $J_3$ term in the energy Eq. (\ref{Hc}) has the same effect as a perpendicular magnetic anisotropy (PMA) term $K S_{z}^2$ used in experiments. Topological solitons have been studied in Ir/Co/Pt  nano-disc multilayered systems, with  the PMA term  $K$ varying spatially  over each layer, with  a linear dependence \cite{kent}. Our results  suggest that layers with a circularly symmetric  inverse square dependence $1/\rho^2$ of the inhomogeneity in the anisotropy $\widetilde{J}_{3}$ will lead to stable hopfion vortices  with a range of $H$ values. Some suggestions with regard to fabricating inhomogeneous magnetic materials have been made in \cite{balakrishnan2}.   
  
  Finally, we note that the inverse  square interaction of our model  that has led to exact solvability  is  reminiscent of a similar interaction between  particles in the well known Calogero-Moser model  which is known to be completely integrable, with connections to diverse fields \cite{CM}. Hence our work  has potential ramifications for other physical systems as well.

We hope that our results will motivate the fabrication of  inhomogeneous, anisotropic  3D magnetic materials that
are  described by  our model, so that  the exact hopfion vortex solutions 
predicted by it can be created in the laboratory and their possible applications in nanotechnology investigated.

\noindent
\textit{Acknowledgments.}{\textbf{\textemdash}} We thank Ayhan Duzgun for help with the figures. The work 
of A.S.  at Los Alamos National Laboratory was carried out under the auspices of the U.S. DOE and NNSA under Contract No. DEAC52-06NA25396.


\begin{thebibliography}{99}
\bibitem{fn} We note here that the lower bound of the energy has  a sublinear behavior in the Faddeev-Skyrme  model as well; however it goes as $H^{3/4}$ \cite{shnir} in this  model which supports hopfions rather than hopfion vortices.
\bibitem{fn2} It must be noted that  Derrick's  scaling  \cite{derrick} is  always applied to the given energy expression of the system, and that the general analysis  is {\em independent}  of the specific  boundary conditions that may be subsequently  used to obtain explicit  solutions supported by it.
\end{thebibliography}
\end{document}